\documentclass[10pt,twoside]{classe-hy_e}
\usepackage{amssymb,amsbsy,amsmath,amsfonts,amssymb,amscd}
\usepackage{latexsym,euscript,exscale,epic,eepic,epsfig}
\usepackage[english,francais]{babel}
\usepackage{times}
\usepackage{rotating}
\input utile.def
\newcommand{\lsim}   {\mathrel{\mathop{\kern 0pt \rlap
  {\raise.2ex\hbox{$<$}}}
  \lower.9ex\hbox{\kern-.190em $\sim$}}}
\newcommand{\gsim}   {\mathrel{\mathop{\kern 0pt \rlap
  {\raise.2ex\hbox{$>$}}}
  \lower.9ex\hbox{\kern-.190em $\sim$}}}
\def\be{\begin{equation}}
\def\ee{\end{equation}}
\def\ba{\begin{eqnarray}}
\def\ea{\end{eqnarray}}
\def\ap{\approx}

\def\snn{\sigma_{\nu N}}

\ComParit{S1296-2147}
\PIT{FLA}
\PXHY{????}
\Add{?}
\Volume{}
\Year{2004}
\FirstPage{}
\LastPage{}
\AuteurCourant{M.~Kachelrie\ss}
\TitreCourant{Particle physics and the UHECR puzzle} 
\Journal
\Rubrique{Physique - Astrophysique}{Physics - Astrophysics}
\SousRubrique{Preprint}{MPP-2004-1}
\PresentePar{}{}
\Recu{}{}{}
\motscles{.../ .../  }
\begin{document}
\selectlanguage{english}
\unitlength1cm
\TitleOfDossier{Ultimate energy particles in the universe}
\TitreDeDossier{Particules d'\'energies ultimes dans l'univers}
\title{%
Status of particle physics solutions to the UHECR puzzle
}
\author{%
Michael Kachelrie\ss~$^{\text{a}}$
}
\address{%
\begin{itemize}\labelsep=2mm\leftskip=-5mm
\item[$^{\text{a}}$]
Max-Planck-Institut f\"ur Physik (Werner-Heisenberg-Institut),
M\"unchen\\
E-mail: mika@mppmu.mpg.de
\end{itemize}
}
\maketitle
\thispagestyle{empty}
\begin{Abstract}{%
The status of solutions to the ultra-high energy cosmic ray
puzzle that involve particle physics beyond the standard model
is reviewed. Signatures and experimental constraints are discussed for
most proposals like the $Z$ burst model and topological
defects (both allowed only as subdominant contribution), supermassive dark
matter (no positive evidence from its key signatures galactic anisotropy 
and photon dominance), strongly interacting neutrinos
or new primaries (no viable models known), and violation of
Lorentz invariance (possible). 
}\end{Abstract}
\selectlanguage{french}
\begin{Ftitle}{%
Solutions \`a l'\'enigme des rayons cosmiques ultra-\'energ\'etiques en physique des particules.
}\end{Ftitle}
\begin{Resume}{%
Nous pr\'esentons une revue des solutions propos\'ees en r\'eponse \`a
l'\'enigme des rayons cosmiques ultra-\'energ\'etiques, faisant
intervenir la physique des particules au-del\`a du Mod\`ele
Standard. Nous r\'esumons les signatures et les contraintes
exp\'erimentales pour la plupart de ces mod\`eles tels que~: la
d\'esint\'egration du $Z$ et les d\'efauts topologiques (tous deux
envisageables seulement en tant que mod\`eles sub-dominants),
mati\`ere noire supermassive (qu'aucune indication, telle
l'anisotropie galactique et production dominante de photons, ne
favorise), les neutrinos \`a interaction forte ou des particules
nouvelles (pas de mod\`eles viables connus), et la violation de
l'invariance de Lorentz (viable). 
}\end{Resume}

\par\medskip\centerline{\rule{2cm}{0.2mm}}\medskip
\setcounter{section}{0}
\selectlanguage{english}

\section{Introduction}

Cosmic Rays are observed in an energy range extending over more than 
eleven decades, starting from subGeV energies up to $3\times 10^{20}$~eV. 
Apart from the highest energies, these particles are thought to be
accelerated in 
our Galaxy, most probably by supernova remnants. Since the galactic
magnetic field cannot isotropize particles with energies
higher than $\sim Z\times 10^{19}$~eV but the arrival directions of
ultra-high energy cosmic rays (UHECRs) are isotropic on large scales,
it is natural to think that UHECRs have an extragalactic
origin. Moreover, the acceleration of protons or nuclei up to
2--3$\times 10^{20}$~eV is difficult to explain with the known
astrophysical galactic sources~\cite{acc}.

\begin{figure}
\epsfig{file=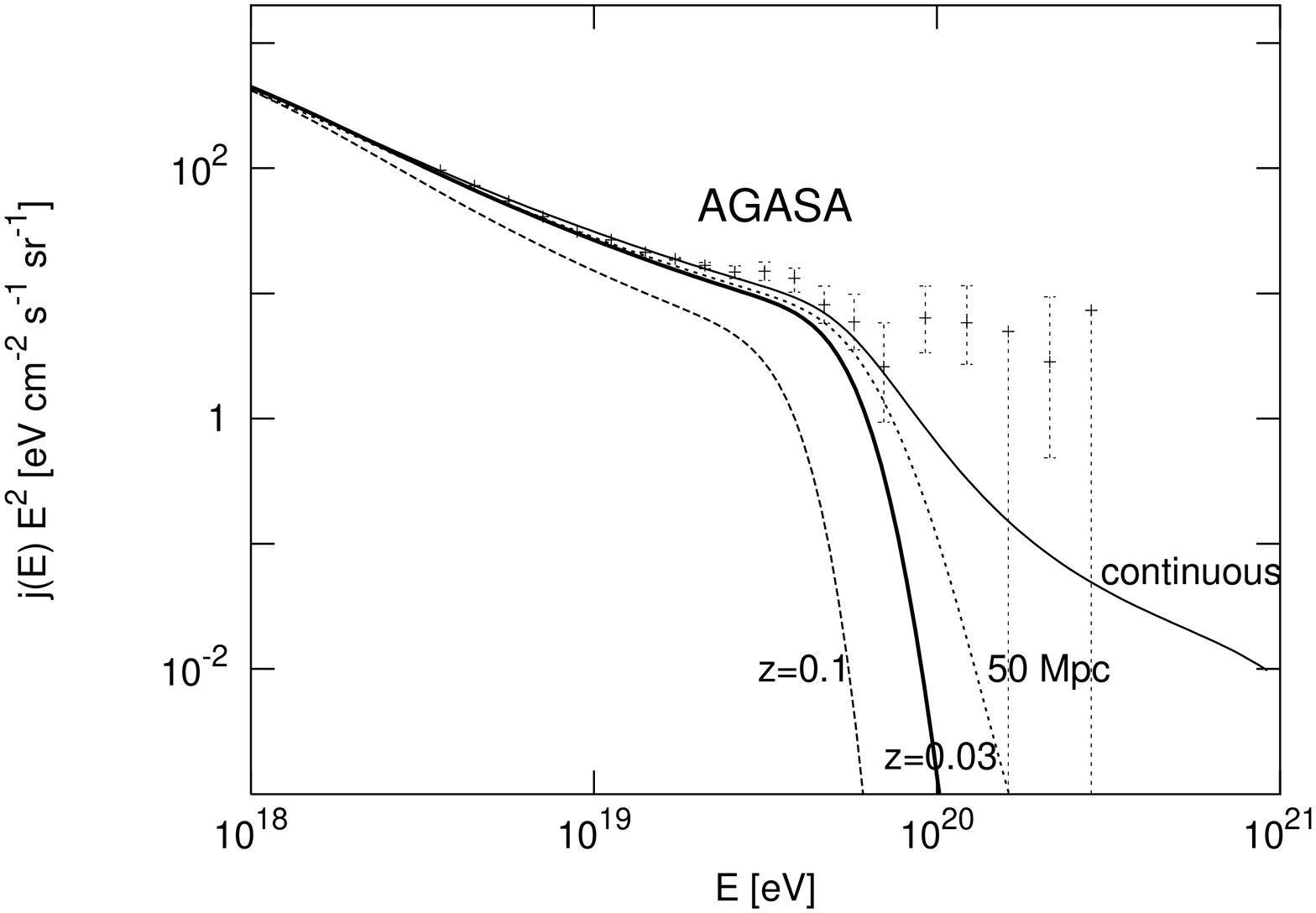,height=7.5cm,angle=270}
\epsfig{file=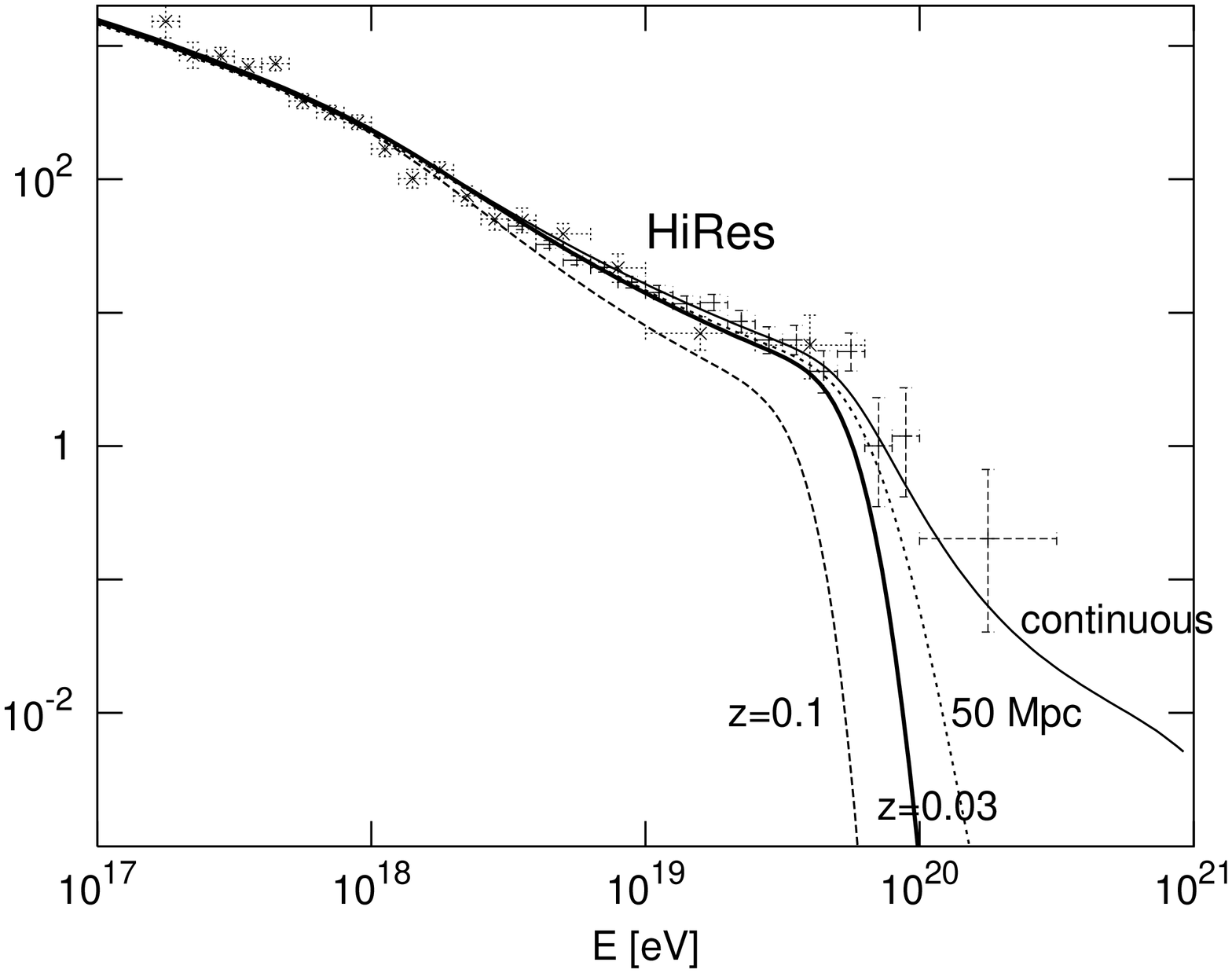,height=7.5cm,angle=270}
\caption{
  Energy spectrum multiplied by $E^2$ with error bars as observed by
  AGASA (left) and 
  HiRes (right) together with the spectrum expected from uniformly
  distributed proton sources with generation spectrum $\propto
  E^{-2.7}$, maximal energy $E_{\rm max}=10^{21}$~eV, and minimal
  distance of the sources as indicated  
  (from Ref.~\cite{Kachelriess:2003yy}).
\label{spec}}
\end{figure}

{\em  Energy spectrum and propagation effects:\/}
The most prominent signature of extragalactic UHECR is the so called
Greisen--Zatsepin--Kuzmin (GZK) cutoff~\cite{GZK}: the energy losses
of protons increase sharply at $E_{\rm GZK} \ap 5\times 10^{19}$~eV, 
since pion-production on cosmic microwave background (CMB) photons, 
$p+\gamma_{3K}\to\Delta^\ast\to N+\pi$, reduces their mean free path
by more than two orders of magnitude compared to lower energies. 
Nuclei exhibit an even more pronounced cutoff at a somewhat higher
energy, while photons are absorbed over a few Mpc due to pair-production
on the radio background. Thus, the UHECR spectrum should dramatically
steepen above $E_{\rm GZK}$ for {\em any\/} homogeneous distribution
of proton or nuclei sources, for more details see Ref.~\cite{prop}.  
How pronounced the GZK cutoff is depends not only on the
shape and maximal energy of the injection spectrum but also on the total
number $N_s$ of sources: The average distance to the nearest sources
increases for decreasing $N_s$, and the GZK cutoff becomes thus more
pronounced. The spectrum shown usually corresponds to a continuous
distribution of sources, i.e. to the limit $N_s\to\infty$, and hence
underestimates the GZK suppression. In Fig.~\ref{spec}, the data from
the two experiments with the currently largest exposure,
AGASA~\cite{AGASA} and HiRes~\cite{HiRes}, are compared to the
expectation for uniformly distributed proton sources with different
minimal distance to the sources~\cite{Kachelriess:2003yy}.  In
particular the flux above $E\gsim 10^{20}$~eV depends strongly on the
used minimal distance. While the AGASA experiment has
detected a significant excess of events above $10^{20}$~eV compared
to the prediction for a continuous source distribution, the flux
measured by HiRes is consistent with this assumption. But if there are
no sources able to accelerate to the highest energies within less than
50~Mpc, then even the low flux observed by HiRes is difficult to
explain. 

Another important consequence of the scattering of UHE primaries,
especially of photons, on background photons is the cascade or EGRET
limit~\cite{baksan}: 
High energy electrons and photons scattering on CMB photons initiate
electromagnetic cascades until their energy is accumulated as
gamma radiation in the MeV--GeV region. The observation of this 
diffuse background by the EGRET experiment~\cite{egret} limits thereby 
the injection of UHE particles. Following the new calculation of the
Galactic foreground of Ref.~\cite{new-egret}, the limit
$\omega_{\rm cas} \leq \omega_{\rm obs} = 2\times 10^{-6}$~eV/cm$^3$ 
results for any diffuse injection of
electromagnetic energy during the history of the Universe. As we will
see later, this is a severe constraint for the $Z$ burst and most
topological defect models.

{\em Arrival directions and clustering~\cite{clay}:\/}
No significant enhancement of the arrival directions of the UHECRs above
$4\times 10^{19}$~eV  towards the galactic or supergalactic plane is
found, their arrival directions are scattered isotropically on scales
larger than 5 degrees. However, about 20\% of the events are clustered
in angular doublets or even triplets; both triplets are found near the
supergalactic plane. The chance probability to observe the
clustered events in the case of an isotropic distribution of arrival
directions was estimated to be $<1\%$~\cite{Uc00}.  
Reference~\cite{Burgett:2003yg} 
pointed out that clustering in the AGASA data is
seen only for $E<6\times 10^{19}$~eV, while above this energy the
arrival directions are consistent with the expectation for an
isotropic distribution. This could be, together with the shape of the
AGASA energy spectrum, a hint for a new component in the UHECR
flux above $E\gsim 6\times 10^{19}$~eV.

There are two possible interpretations if clustering is confirmed by
future experiments. 
Either the extragalactic magnetic fields are small, UHE protons are
propagating nearly undeflected and experiments start to see several
events of the same point source, or the extragalactic magnetic fields
are close to the upper bound from 
Faraday rotation measurements. In the latter case, deflections would
prevent the identification of few sources nearby and magnetic
lensing can be used to explain clustering. Experimentally, the
two options can be distinguished by an (auto-) correlation analysis of
the arrival directions. Firstly, the shape of the angular
autocorrelation function is determined for point sources by the
angular resolution function of the experiment, while a
broader peak around zero is expected for magnetic lensing.
Although the significance of the autocorrelation in the AGASA data
is maximal choosing as bin size $2.5^\circ$, i.e. the angular
resolution of AGASA, the data set is too sparse to disfavour thereby
clearly magnetic lensing. 
Secondly, a correlation analysis of the UHECR arrival directions with
possible candidate sources should reveal their true sources if the
deflection through magnetic fields is small enough  (see below).

Theoretically, the magnitude and structure of extragalactic magnetic
fields is rather uncertain. On one hand, a recent constrained
simulation~\cite{Dolag:2003ra} of large-scale structures favors small
extragalactic magnetic fields. The deflection of a proton with, e.g,
$E=4\times 10^{19}$~eV found in~\cite{Dolag:2003ra} is less than
$2.5^\circ$ in 95\% (70\%) of the sky for a propagation distance of
100~Mpc (500~Mpc).  
Reference~\cite{ems}, on the other hand, points out that sources tend
to sit in regions of high density and strong magnetic fields, an
effect not taken into account in Ref.~\cite{Dolag:2003ra}. As a
consequence, the deflection angles found in \cite{ems} are much larger,
implying that source identification may be not possible in their
scenario.

The total number $N_s$ of UHECR sources, i.e. including those not
detected yet, can be determined by the fraction of clustered
events~\cite{Wa97}. As $N_s$ decreases, the sources have to become
brighter for a fixed UHECR flux and therefore the probability for
clustering increases. The analysis of Ref.~\cite{Du00}, assuming small
magnetic fields, showed that
$\sim 400$ sources of cosmic rays with $E>10^{20}$~eV should be inside
the GZK volume, compared to $\sim 10$~GRB sources or $\sim 250$~AGNs
of which only a small fraction is thought to be UHECR sources. However, 
the statistical uncertainties of this analysis are very large,
because of the small number of events above $E=10^{20}$~eV.   
References~\cite{pmks} found for the density $n_s$ of
uniformly distributed sources using AGASA events above $4\times
10^{19}$~eV as best-fit value $n_s\sim 10^{-5}/$~Mpc$^3$, i.e. a value
compatibel with the density of AGNs.

{\em Correlations:\/}
Tinyakov and Tkachev found a significant, but currently disputed
correlation of UHECR arrival directions with BL
Lacs~\cite{bllac}. The BL Lacs which correlate with the UHECRs 
are located at very large (redshift $z\sim 0.1$) or unknown distances.
If it can be shown with an increased data set of UHECRs that this
correlation holds at energies $E\gsim 6\times 10^{19}$~eV, then
protons that cannot reach us from these distances cannot
explain the UHECR data.

The difficulty to accelerate particles in astrophysical accelerators
up to energies $E\gsim 10^{20}$~eV, the extension of the UHECR spectrum
beyond the  GZK cutoff, the missing correlation of the UHECR arrival
directions with powerful nearby sources and, more recently, their
possible correlation with BL Lacs has prompted many proposals to
explain this puzzle that involve particle physics beyond the standard
model (SM). In the
next sections, the most prominent ones will be discussed and their 
current status will be reviewed.

\section{Neutrinos as primaries or messenger particles}
Neutrinos are the only known stable particles that can traverse
extragalactic space without attenuation even at energies $E\gsim
E_{\rm GZK}$, thus avoiding the GZK cutoff. Therefore, it has been
speculated that the UHE primaries initiating the observed air showers
are not protons, nuclei or photons but neutrinos~\cite{alt,mittel}. 
However, neutrinos are in the SM deeply
penetrating particles  producing mainly horizontal not vertical
extensive air showers (EAS).
Therefore, either one has to postulate new interactions that enhance
the UHE neutrino-nucleon cross section by a factor $\sim 10^6$ or 
neutrinos have to be converted ``locally'' into hadrons or photons. 
 
\subsection{Annihilations on relic neutrinos -- $Z$ burst model}
In the later scheme~\cite{We99/Fa99}, UHE neutrinos from distant sources
annihilate with relic neutrinos on the $Z$ resonance. The
fragmentation products from nearby $Z$ decays, i.e. mainly photons,
are supposed to be the primaries responsible for the EAS above the GZK
cutoff.  For energies of the primary neutrino of
$E_0\sim 4\times 10^{22}$~eV, the mass of the relic neutrino should be
$m_\nu = m_Z^2/(2E_0)\sim 0.1$~eV to scatter resonantely, a
value compatible with atmospheric neutrino oscillation data. There
are, however, severe constraints on this model:

Primary protons have to be accelerated to extremely high energies, 
$E\gsim 10^{23}$~eV, in order to produce on a beam-dump in 
astrophysical sources via $p+\gamma\to$~all or $p+p\to$~all UHE
neutrinos as secondaries. The photons which are unavoidably produced
in the same reactions have to be hidden inside the source, otherwise
the diffuse MeV--GeV photon background is overproduced. 
No astrophysical accelerator of this kind is known. Moreover, hidden
sources are unable to produce large fluxes of high-energy neutrinos
above that energy at which the decay length of charged pions becomes
equal to its scattering length. 
Another problem is the extreme luminosity of the astrophysical sources 
needed in this model~\cite{Berezinsky:2002hq}: From the required flux
of resonant neutrinos $I_{\nu}(E_0)$ one can estimate the neutrino
energy density $\omega_{\nu}$ as  
$\omega_{\nu} \approx (2.4 - 3.6)\times 10^{-13} m_{\rm eV}^{-0.5}$~erg/cm$^3$.
The resulting neutrino  luminosity of a source, 
$L_{\nu} \sim \omega_{\nu}/(n_st_0)$, where $n_s$ is the source density 
and $t_0$ the age of the Universe, is unacceptably high:
$(8 -12)\times 10^{44}$~erg/s, if the sources are normal galaxies, and 
$(8 -12)\times 10^{46}$~erg/s in the case of Seyfert galaxies.

As possible way-out, the authors of Ref.~\cite{Gelmini:1999ds}
combined the $Z$ burst model and superheavy dark matter (SHDM): they
suggested that SHDM particles decay 
exclusively to neutrinos thereby avoiding both the acceleration
problem and photon production in astrophysical sources. However, 
higher-order corrections to the tree-level process
$X\to\bar\nu\nu$ give rise to an electroweak cascade transferring
around 20\% of the initial energy to photons and
electrons~\cite{Berezinsky:2002hq}. Thus the EGRET limit can be applied
also to this variant of the $Z$ burst model.

A combination of the WMAP observations of the CMBR fluctuations and 
the 2dFGRS galaxy count limits the sum of all
neutrino masses as $\sum_i m_{\nu_i}\lsim 1.0$~eV at 95\%~CL
(cf., e.g., Ref.~\cite{steen}). For such small masses, the overdensity
$\delta$ of neutrinos in our Local Group of galaxies is also small,
$\delta\lsim 10$, on a length scale of $1~$Mpc~\cite{singh}. Therefore
one expects a rather pronounced GZK cutoff, needs very large
neutrino fluxes and has problems with the cascade limit. 
The latter point can be understood from a simple estimate: The energy
density dumped into electromagnetic cascades by the $Z$ burst
mechanism during the life-time of the Universe is 
%
%
\begin{equation}
 \omega_{\rm cas} \sim
 \frac{1}{2} f_\pi E_0 \dot n_Z t_0 \,, 
\end{equation}
where $f_\pi\sim 0.7$ is the branching ratio of $Z$ decays into pions
and $\dot n_Z$ is the rate of $\bar\nu\nu\to Z$ scatterings. The
cascade limit, $\omega_{\rm cas}\leq \omega_{\rm obs}$, translates
into a bound on $\dot n_Z$, and therefore, on the photon flux 
\begin{equation}
 I_\gamma(E) = 
 \frac{1}{4\pi} \dot n_Z R_\gamma(E) D_\gamma(x)/E_0 \leq 
 \frac{\omega_{\rm obs}}{2\pi f_\pi E_0^2 t_0} \: R_\gamma(E) D_\gamma(x) \,.
\end{equation}
Here, $R_\gamma(E)$ is the attenuation length of photons and
$D_\gamma(x)$ with $x=2E/m_Z$ is the differential energy spectrum per
$Z$ decay. Estimating $I_\gamma(E)$ at $E'=10^{20}$~eV and inserting
$R_\gamma(E')\sim 10^{25}$~cm and $D(x')\sim 20$ results in the bound 
$I_\gamma(E')\lsim 10^{-33}$~sr$^{-1}$~GeV$^{-1}$~cm$^{-2}$~s$^{-1}$, 
while the observed UHECR flux is 1--2 orders of magnitude higher.

\begin{figure}
\begin{center}
\epsfig{file=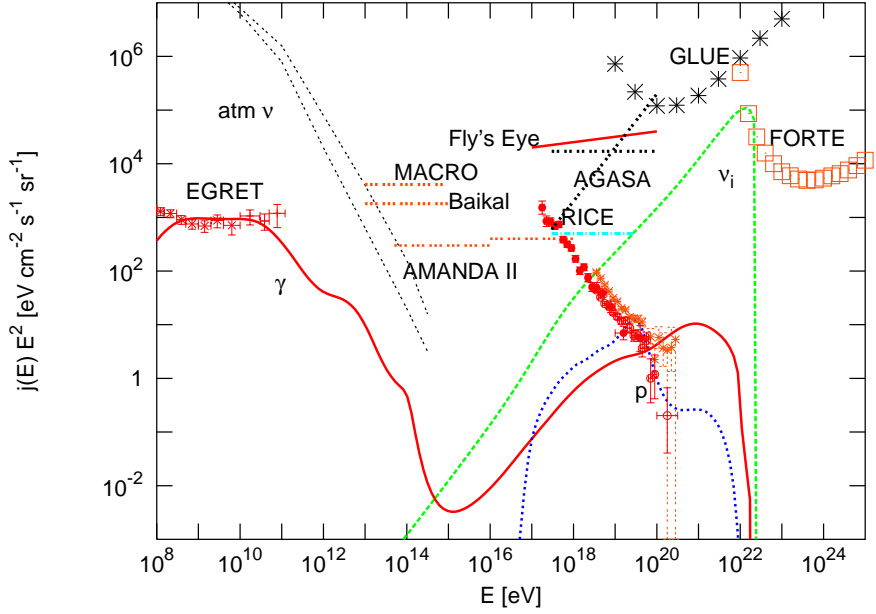,height=12cm,angle=270}
\end{center}
\caption{
Expected fluxes in the $Z$ burst model for an optimal choice of free
parameters together with various limits for UHE neutrinos fluxes and the new
EGRET limit; from Ref.~\cite{ss}.
\label{z}}
\end{figure}

This discrepancy is weakened assuming e.g. that the $Z$ burst
mechanism started operation only recently and by a more accurate analysis.
Reference~\cite{ss} calculated numerically the expected fluxes in the
$Z$ burst model and compared them to the improved
limit~\cite{new-egret} from EGRET and new experimental limits on the
UHE neutrino flux from FORTE~\cite{forte} and GLUE~\cite{glue}.
Their results are shown in Fig.~\ref{z} for $m_\nu=0.33$~eV, the
unrealistic case of an only neutrino emitting source, and an optimal
choice of free parameters; for all other cases, the conflict is more
severe.

\subsection{Strongly interacting neutrinos}
Most models introducing new physics at a scale $M$ to produce large
cross sections for UHE neutrinos fail because experiments generally
constrain $M$ to be larger than the weak scale, $M\gsim m_Z$, and
unitarity limits cross sections to be $O(\sigma_{\rm tot})\lsim 1/M^2
\lsim 1/m_Z^2$.  String theories with large extra
dimensions \cite{ex} are different in this respect: If the
SM particles are confined to the usual (3+1)-dimensional space and
only gravity propagates in the higher dimensions, the
compactification radius $R$ of the large extra dimensions can be
large, corresponding to a {\em small\/} scale $1/R$ of new physics. 
From a four-dimensional point of view the higher dimensional graviton
in these theories appears as an infinite tower of Kaluza-Klein (KK)
excitations with mass squared $m_{n}^2={n}^2/R^2$.  Since
the weakness of the gravitational interaction is partially compensated
by the large number of KK states and cross sections of reactions
mediated by spin 2 particles are increasing rapidly with energy, it
has been argued in Refs.~\cite{mittel} that neutrinos could
initiate the observed vertical showers at the highest energies.
However, the naively found growth of $\snn\propto s^2$ violates
unitarity and an unitarization procedure has to applied. The
unitarized cross section is roughly three orders of magnitude too
small, and also the energy transferred in each interaction is not
sufficient to explain the observed properties of EAS~\cite{no-large}.
For small enough impact parameters in the neutrino-nucleon collision, black
hole (BH) production becomes important~\cite{BH}. Using in a simplistic
picture a geometric cross section for BH production,
$\sigma_{\rm BH}\sim \pi R_S^2$ where $R_S$ is the Schwarzschild
radius of a BH with mass equal to the
center-of-mass energy of the collision on the parton level, the cross
section has roughly the same size as the one for KK scattering and is
thus also too small~\cite{BH-CR}.

More recently, Ref.~\cite{FKRT} speculated that the neutrino-nucleon
cross section above $E\sim 10^{18}$~eV is enhanced by a factor 
$\sim 10^5$ by 
non-perturbative electroweak instanton contributions. The numerical
calculations of Ref.~\cite{Bezrukov:2003qm} found that instanton
induced processes are much more heavily suppressed than suggested
by~\cite{FKRT}.  
However, it is instructive to ask if strongly interacting neutrinos
can mimic at all  in this model extensive air showers initiated by
protons. 
At $E\leq 10^{20}$~eV, the cross section is bounded by 
$\sigma_{\nu p}\leq 3$~mbarn~\cite{bound}. 
Thus the difference between $\sigma_{\nu p}$ and 
$\sigma_{pp}^{\rm inel}$ is still large even at UHE in this model, 
$\sigma_{pp}^{\rm inel}/\sigma_{\nu p}\sim 40$, but it
becomes smaller considering the scattering on air nuclei: While
scattering of protons on nuclei with $A>1$ is already close to the
black disc limit ($\sigma_{pA}^{\rm inel}\propto A^{2/3}$), it is
reasonable to assume no shadowing, $\sigma_{\nu A}\propto A$, for
neutrino-nucleus scattering. Even so, the development of a neutrino
induced shower is considerably delayed having its shower maximum around 
$\geq 1400$~g/cm$^2$. 
Experiments as HiRes or the Pierre Auger Observatory that are able to
measure the whole shower development of an EAS in fluorescent light
should clearly see this difference. 

In summary, experimental and theoretical constraints make it very
unlikely that neutrinos can explain---either as messenger particles or as
primaries---the observed vertical EAS. Nevertheless, both cases offer 
future experiments exciting possibilities: the discovery of the relic
neutrino background and, perhaps, a measurement of the absolute
neutrino masses via the $Z$ burst mechanism, or the discovery of new
contributions to the neutrino-nucleon interaction in horizontal
EAS~\cite{BH-CR,nu-disc}.

\section{Top--down models}
Top--down model is a generic name for all proposals in which the
observed UHECR primaries are produced as decay products of some
superheavy particles $X$ with mass $m_X\gsim 10^{12}$~GeV. 
These $X$ particles can be either metastable
or be emitted by topological defects at the present epoch.

\begin{figure}
\begin{center}
\epsfig{file=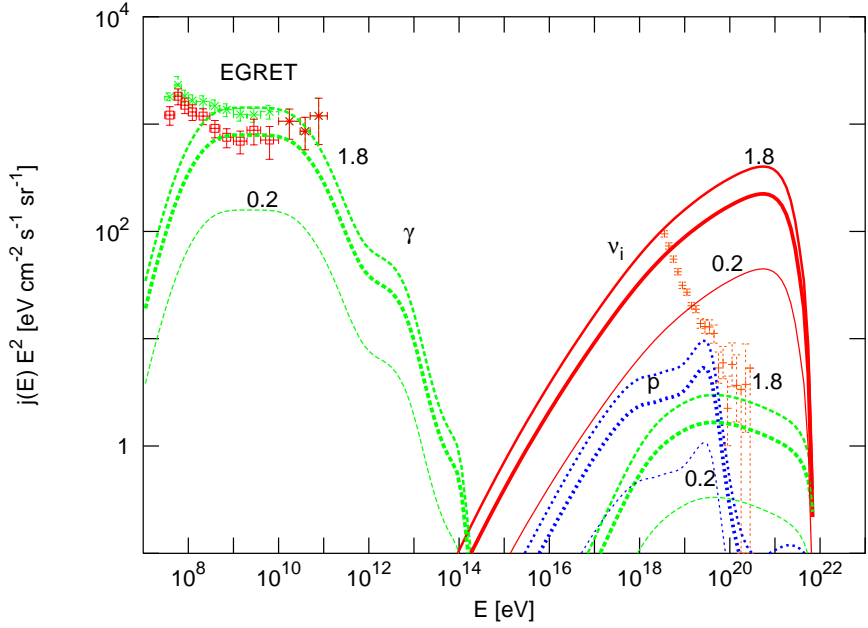,height=12cm,angle=270}
\end{center}
\caption{
Proton, photon and neutrino fluxes in a TD model with $M_X=2\times
10^{13}$~GeV, evolution $\dot n_X \propto t^{-3}$ and continuous
distribution of sources. The fraction of the MeV--GeV diffuse photon
background MeV--GeV contributed by these sources is chosen as 0.2,
1, and 1.8; from Ref.~\cite{ss}.
\label{td}}
\end{figure}

\subsection{Topological defects}
Topological defects (TDs)~\cite{td} such as (superconducting)
cosmic strings,  monopoles, and hybrid defects can be
effectively produced in non-thermal phase transitions during the
preheating stage~\cite{Kh98/Ku99}. Therefore the presence of TDs
is not in conflict with an inflationary period of the early
Universe. They can naturally produce particles with high enough
energies but have problems to produce large enough fluxes of UHE
primaries.

{\em Ordinary strings\/} can produce UHE particles e.g. when string loops
self-intersect or when two cusp segments overlap and annihilate. In
the latter case, the maximal energy of the produced fragmentation
products is not $m_X/2$ but can be much larger due to the high
Lorentz factors of the ejected $X$ particles.

{\em Superconducting strings:\/}
Cosmic strings can be superconducting in a broad class
of particle models. Electric currents can be induced in
the string either by a primordial magnetic field that decreases during
the expansion of the Universe or when the string moves through
galactic fields at present. If the current reaches the vacuum
expectation value of the
Higgs field breaking the extra U(1), the trapped particles are ejected
and can decay.

{\em Monopolium M,\/} a bound-state of a monopole--antimonopole pair, was
the first TD proposed as UHECR source~\cite{Hi83}. It clusters like
Cold Dark Matter (CDM) and is therefore an example for SHDM. The
galactic density of monopoles is constrained by the Parker limit: the
galactic magnetic field should not be eliminated by the acceleration of
monopoles. Reference~\cite{Bl99} concluded that the resulting limit on
the UHECR flux produced by  Monopolium annihilations is 10 orders of
magnitude too low.

{\em Cosmic necklaces\/} are hybrid defects
consisting of monopoles connected by a string. These defects are produced 
by the symmetry breaking $G\to H\times U(1) \to H\times Z_2$, where
$G$ is semi-simple.
In the first phase transition at scale $\eta_m$, monopoles are 
produced. At the second phase transition, at scale  $\eta_s<\eta_m$, each 
monopole gets attached to two strings. 
The basic parameter for the evolution of necklaces is the ratio
$r=m/(\mu d)$ of the monopole mass $m$ and the mass of the string 
between two monopoles, $\mu d$, where $\mu \sim \eta_s^2$ is the mass 
density of the string and $d$ the distance between two monopoles.
Strings lose their energy and contract through gravitational radiation.
As a result, all monopoles annihilate in the end producing
$X$ particles. Reference~\cite{Be97} argued that for a reasonable range of
parameters the model predicts a UHECR flux close to the observed one.
A numerical study~\cite{Si00} of the evolution of necklaces
found that the lifetime of necklaces is generally much shorter than
the age of the Universe. An exception is the case
$\eta_m\gg\eta_s\sim 100$~GeV~\cite{Bl99}.

The main observational constraint for topological defect models 
is the EGRET limit.    
Another general reason for the low fluxes is the large distance
between TDs. Then the
flux of UHE particles is either exponentially suppressed or strongly
anisotropic if a TD is nearby by chance. An exception is the necklace
model where the distance $\sim t_0/\sqrt{r}$ between necklaces can be
as small as 10~kpc. Therefore we discuss in the following only this model.

The rate of $X$ particle production by necklaces at time $t$ can be
estimated as~\cite{Be97}
\begin{equation}
\dot{n}_X \sim \frac{r^2 \mu}{t^3 M_X} 
\label{rate}
\end{equation}
and the resulting cascade energy density is given by
\begin{equation}
\omega_{\rm cas}=\frac{1}{2}f_{\pi}r^2\mu \int_0^{t_0} \frac{dt}{t^3}
\frac{1}{(1+z)^4}=\frac{3}{4}f_{\pi}r^2\frac{\mu}{t_0^2} \,,
\label{omega}
\end{equation}
where $f_{\pi}\sim 1$ is the fraction of the total energy transferred
to the cascade. Using the bound on $\omega_{\rm cas}$ and
$t_0=13.7$~Gyr, the limit $r^2\mu \leq 8.9\times 10^{27}$~GeV$^2$
follows~\cite{Aloisio:2003xj}.   

\begin{figure}
\begin{picture}(16,6)
\put (-0.1,1.5) 
{\begin{sideways}
                   {$E^3 J(E)/$m$^{-2}$s$^{-1}$eV$^{2}$}
                \end{sideways}}
\put (4,0.) {$E$/eV}
\put (1.6,2.8) {$\nu$}
\put (1.6,1.25) {p}
\put (3.3,1.6) {$\gamma$}
\put (5.8,3.6) {p+$\gamma$}
\put (0.3,5.5) {
\epsfig{file=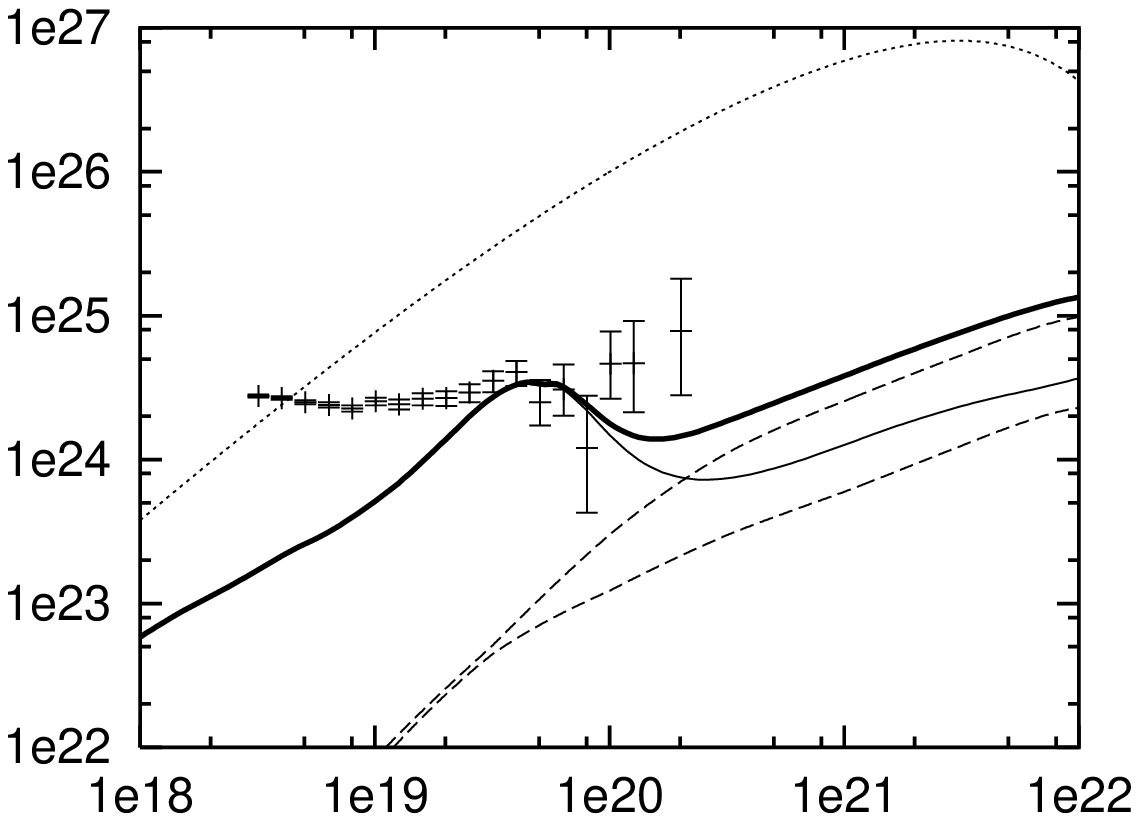,height=7.5cm,angle=270}
}
\put (11.5,0.) {$E$/eV}
\put (7.4,5.7) {
\epsfig{file=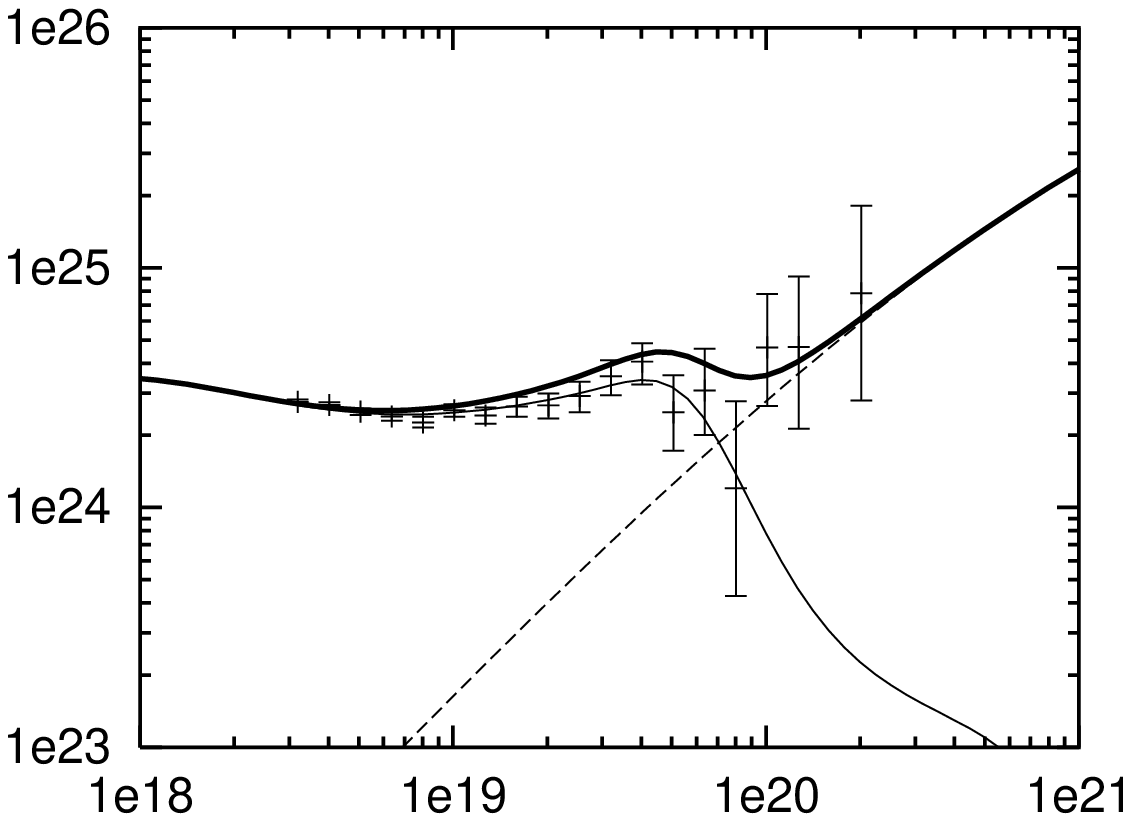,height=8.cm,angle=270}
}
\end{picture}
\caption{
a.
Diffuse spectra from necklaces together with the AGASA data: photon
fluxes are shown for 
two cases of absorption, the thick continuous curve gives the sum
of the proton and the higher photon flux.
4.b:
Comparison of the UHECR flux in the SHDM model with the AGASA
data, photons from SHDM decays (dashed line), spectrum of extragalactic
protons (thin solid line) in the non-evolutionary model of
Ref.~\cite{BGG03} and the sum of these two spectra shown by the thick curve; 
both figures from Ref.~\cite{Aloisio:2003xj}.  
\label{ABK}}
\end{figure}

In Fig.~\ref{ABK}.a, the diffuse fluxes in the necklace model are shown
for  $r^2\mu = 4.7\times 10^{27}$~GeV$^2$, i.e. roughly a factor two
below the cascade bound, and $M_X= 1\times 10^{14}$~GeV. In contrast
to earlier calculations using the MLLA (SUSY) QCD fragmentation
functions, the flux in the necklace model for UHECR is now below the flux
measured by AGASA at $E\geq 10^{20}$~eV. This is the consequence  
of the steeper fragmentation spectra of $X$ particles found in
Refs.~\cite{Be00,Aloisio:2003xj} and used in the calculation
of Fig.~\ref{ABK}. Thus UHE particles from necklaces can serve only as
an additional component in the observed UHECR flux.

A similar conclusions was reached in Ref.~\cite{ss} based on
different assumptions:  
Figure~\ref{td} shows their proton, photon and neutrino fluxes for a TD
model with $M_X=2\times 10^{13}$~GeV, injection rate $\dot n_X \propto
t^{-3}$ (as e.g. in the necklace model) and continuous
distribution of sources. The fraction MeV--GeV photons from this model
contribute to the diffuse photon background is varied between 0.2, 1,
and 1.8. In the calculation of Ref.~\cite{ss} the QCD MLLA
fragmentation functions were used. Here,  the new EGRET limit (lower
set of error bars on the left, in red)
is essential and allows only a sub-dominant contribution to the UHECR
flux from necklaces.

\subsection{Superheavy dark matter}
Superheavy metastable relic particles were proposed in
Refs.~\cite{bkv97,kr97} as UHECR source. They constitute
(part of) the CDM and, consequently, their abundance in
the galactic halo is enhanced by a factor $\sim 5\times 10^4$ above
their extragalactic abundance. 
Therefore, the proton and photon flux is dominated by the halo
component and the GZK cutoff is avoided, as was pointed out
in Ref.~\cite{bkv97}. The quotient $r_X=\Omega_X (t_0/\tau_X)$ of relic
abundance $\Omega_X$ and lifetime $\tau_X$ of the $X$ particle is
fixed by the UHECR flux, $r_X\sim 10^{-11}$. The value of $r_X$ is not
predicted in the generic SHDM model, but calculable as soon as a
specific particle physics and cosmological model is fixed.

There exist several plausible non-equilibrium production mechanisms. 
The most promising one is the gravitational production of the $X$
particles by the non-adiabatic change of the scale factor of the
Universe at the end of inflation, during the transition from the
de-Sitter to the radiation dominated phase~\cite{ch/kz}. In this
scenario, the gravitational coupling of the $X$ field to the
background metric yields independent of any specific particle physics
model the present abundance  
$\Omega_X h^2 \sim (M_X/10^{13}{\rm GeV})^2 (10^{9}{\rm GeV}/T_R)$, 
provided that $M_X\lsim H_\ast$. Here, $T_R$ denotes the reheating
temperature of the Universe and $H_\ast\sim 10^{13}$~GeV the effective
Hubble parameter at the end of inflation. Thus, SHDM could constitute the
main component of CDM for a very interesting set of parameters.
Other mechanisms proposed are thermal production during reheating, 
production through inflaton decay at the preheating phase, or through
the decay of hybrid defects.

The lifetime of the superheavy particle has to be in the range
$10^{17}~{\rm s}\lsim \tau_X\lsim 10^{28}$~s, i.e. longer or much
longer than the age of the Universe. Therefore it is an obvious
question to ask if such an extremely small decay rate can be obtained
without fine-tuning. A well-known example of how metastability can be
achieved is the proton: in the standard model B--L is a conserved
global symmetry, 
and the proton can decay only via non-renormalizable operators. 
Similarly, the $X$ particle could be protected by a new global symmetry
which is only broken by higher-dimensional operators suppressed by
$M^{d}$, where for instance $M\sim M_{\rm Pl}$ and $d\geq 7$ is possible. 
The case of discrete gauged symmetries has been studied in detail in
Refs.~\cite{Ha98}. Another possibility is that the global symmetry is
broken only non-perturbatively, either by wormhole \cite{bkv97} or
instanton \cite{kr97} effects. Then an exponential suppression of the
decay process is expected and lifetimes $\tau_X\gsim t_0$ can be
naturally achieved. 

An example of a SHDM particle in a semi-realistic
particle physics model is the crypton~\cite{El90}.
Cryptons are bound-states from a strongly interacting hidden sector of
string/M theory. Their mass is determined by the non-perturbative
dynamics of this sector and, typically, they decay only through
high-dimensional operators. For instance, flipped SU(5) motivated by
string theory contains bound-states with mass $\sim 10^{12}$~GeV and 
$\tau\sim 10^{15}$~yr~\cite{Be99}. Choosing $T_R\sim 10^5$~GeV results
in $r_X\sim 10^{-11}$, i.e. the required value to explain the UHECR
flux above the GZK cutoff. This example shows clearly that the SHDM
model has no generic ``fine-tuning problem.'' Other viable candidates
suggested by string theory were discussed in Ref.~\cite{cfp}.

\subsection{Signatures of top-down models}
Superheavy dark matter  has several clear signatures: 
1. No GZK cutoff, instead a flat spectrum (compared to astrophysical
sources) up to $m_X/2$. 
2. Large neutrino and photon fluxes compared to the proton flux.
3. Galactic anisotropy. 
4. If $R$ parity is conserved, the lightest supersymmetric particle
(LSP) is an additional UHE primary. 
5. Small-scale clustering of the UHECR arrival
directions gives possibly additional constraints.

{\em 1. Spectral shape:\/}
The fragmentation spectra of superheavy particles calculated by
different methods and different groups~\cite{Be00,tb,Aloisio:2003xj}
agree quite well, cf. Ref.~\cite{Aloisio:2003xj} for a detailed
discussion. This allows to consider the spectral shape as a signature
of models with decays or annihilations of superheavy particles. The 
predicted spectrum in the SHDM model, $dN/dE\propto E^{-1.9}$, cannot 
fit the observed UHECR spectrum at energies $E\leq (6$--$8)\times
10^{19}$~eV. Thus only events at $E\geq (6$--$8)\times 10^{19}$~eV, and most 
notably the AGASA excess at these energies, can be explained in this model.
A two-component fit from Ref.~\cite{Aloisio:2003xj} using protons from
uniformly, continuously distributed extragalactic astrophysical sources
and photons from SHDM is shown in Fig.~\ref{ABK} together with the
experimental data from AGASA.

\begin{figure}
\begin{picture}(16,6)
\put (0.,3) 
{\begin{sideways}
                   $\gamma/N$
                   \end{sideways}}
\put (4.,0.25) {$x$}
\put (0.3,5.7) {\epsfig{file=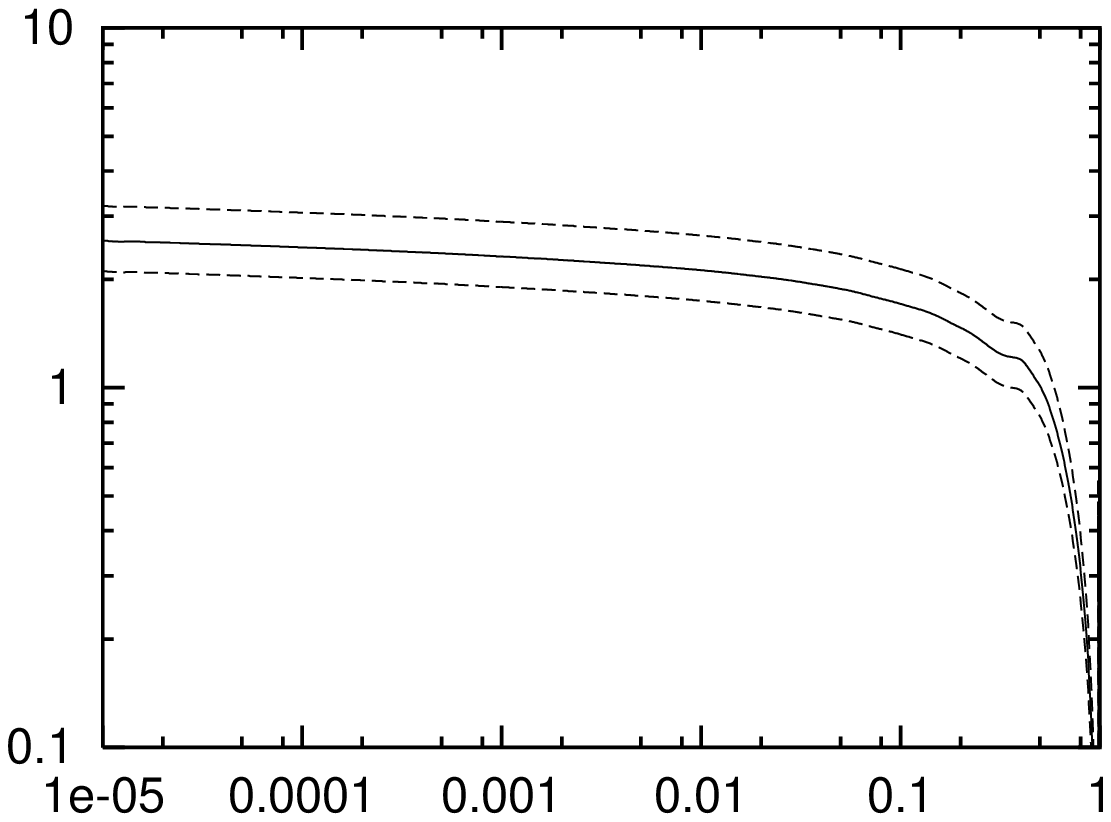,height=7cm,angle=270}}
\put (7.6,3) {\begin{sideways}
                   $N_e$
                   \end{sideways}}
\put (11.4,0.25) {$X_{\max}$, g/cm$^2$}
\put (7.9,0.7) {\epsfig{file=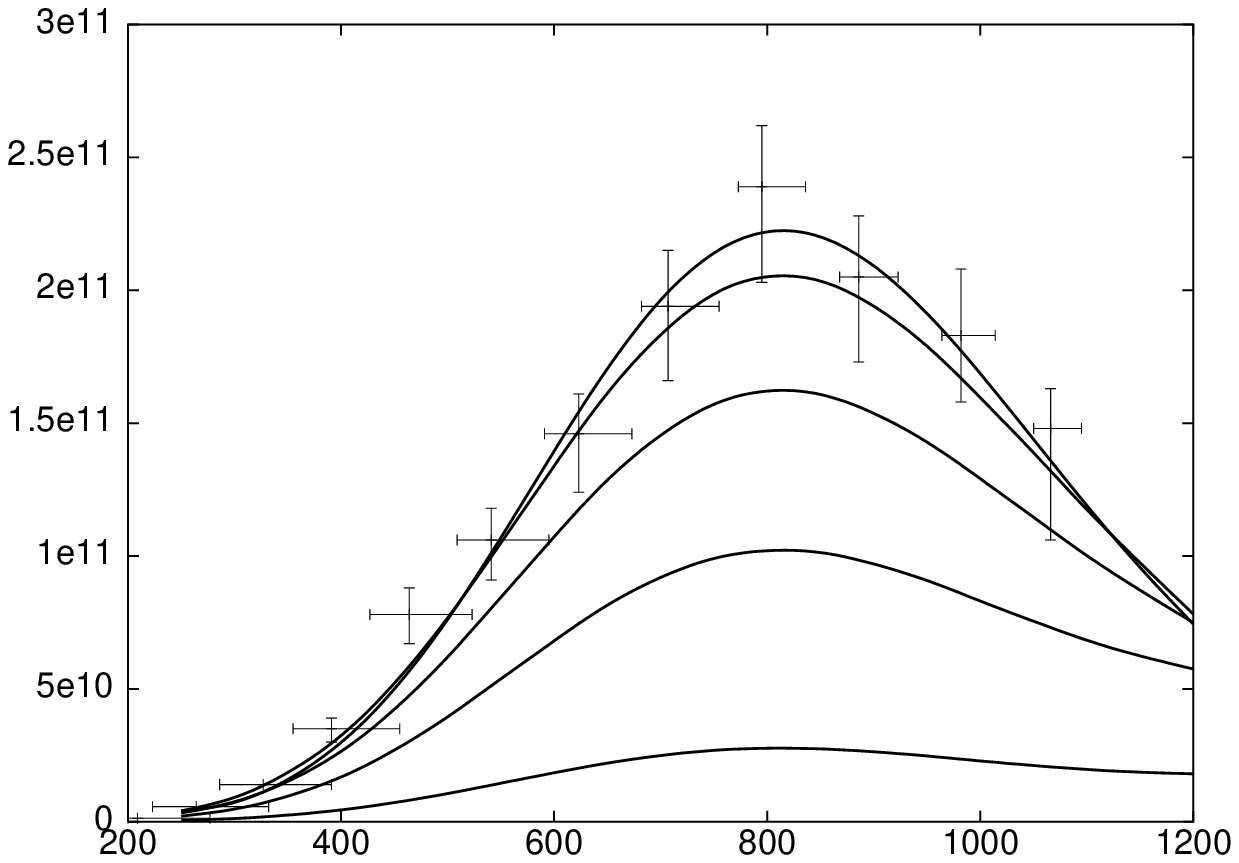,width=7.2cm,angle=0}}
\end{picture}
\caption{
a:
Photon/nucleon ratio as function of $x=2E/M_X$, the band illustrates the
uncertainty due to hadronization; from Ref.~\cite{Aloisio:2003xj}.
5.b:  
Comparison of Flye's Eye highest energy event with the 
longitudinal shower profile for EAS of energy $E_0=3\times 10^{20}$~eV
initiated (from top to down) by protons and by glueballinos with 
$M_{\tilde g}=2,5,10$ and 50~GeV. The shower profiles are shifted so that
their $X_{\max}$ agrees with the observed shower maximum; 
from Ref.~\cite{Berezinsky:2001fy}.
\label{fit}}
\end{figure}

{\em 2. Chemical composition~\cite{watson}:\/}
Since at the end of the QCD cascade quarks combine more easily to
mesons than to baryons, the main component of the UHE flux are
neutrinos and photons from pion decay.
Therefore, a robust prediction of this model is photon dominance with
a photon/nucleon ratio of $\gamma/N \simeq 2$--3, becoming smaller at
the largest $x=2E/M_X$. This ratio is shown in Fig.~\ref{fit}.a as
function of $x$ together with a band illustrating the uncertainty due to
the hadronization process~\cite{Aloisio:2003xj}.

The muon content of photon induced EAS at $E>1\times 10^{20}$~eV is
high, but lower by a factor 5--10 than in hadronic showers~\cite{AhPl}. 
It has been recently measured in a sub-array of AGASA~\cite{AGASA-gamma}. 
From eleven events at $E>1\times 10^{20}$~eV, the muon density was
measured in six. In two of them with 
energies about $1\times 10^{20}$~eV, the muon density is almost twice
higher than predicted for gamma-induced EAS.  The muon content of the
remaining four EAS  marginally agrees with that predicted for
gamma-induced showers. The contribution of extragalactic protons for
these events is negligible, and the fraction of nucleons in the total
flux can be estimated as $0.25 \leq N/{\rm tot}\leq 0.33$. This
fraction gives a considerable contribution to the probability of
observing four showers with slightly increased muon content. Not
restricting severely the SHDM model, the AGASA events give no evidence
in favor of it. 

Reference~\cite{Av00} finds analyzing the Haverah Park data that above
$4\times 10^{19}$~eV less than 55\% of the UHE primaries can be photons.
Since protons from ``normal'' astrophysical sources dominate the flux
up to $(6-8)\times 10^{19}$~eV and the flux is steeply falling with
energy, this result does not constrain the SHDM model.

The Pierre Auger Observatory~\cite{Auger} has great potential to
distinguish between photon 
and proton induced EAS through the simultaneous observation of UHECR
events in fluorescent light and with water Cherenkov detectors: while
for a proton primary both methods should give a consistent
determination of the primary energy, the ground array should
systematically underestimate the energy of a photon primary. Moreover,
the interaction of the photon with the geomagnetic field should induce
an anisotropy in the flux.

{\em 3. Galactic anisotropy:\/} 
The UHECR flux from SHDM should show a galactic
anisotropy~\cite{Du98}, because the Sun is not in the center of the
Galaxy. The degree of this anisotropy depends on how strong the CDM is
concentrated near the galactic center -- a question under debate.
Since experiments in the northern hemisphere do not see the Galactic center,
they are not very  sensitive to a possible anisotropy of arrival
directions of UHECR from SHDM.  In contrast,  the Galactic 
center was visible for the old Australian SUGAR experiment~\cite{SUGAR}.
The compatibility of the SHDM hypothesis with the SUGAR data was
discussed recently in Refs.~\cite{Kachelriess:2003rv,Kim:2003th}. In
Ref.~\cite{Kachelriess:2003rv},  
the expected arrival direction distribution for a two-component energy
spectrum of UHECRs consisting of protons from uniformly distributed,
astrophysical sources and the fragmentation products of SHDM 
calculated in SUSY-QCD  was compared to the data of the SUGAR
experiment using a Kolmogorov-Smirnov test. Depending on the details
of the dark-matter profile and of the composition of the
two-components in the UHECR spectrum, the arrival
directions measured by the SUGAR array have a probability of
$\sim 5$--20\% to be consistent with the SHDM model.
Also in the case of the galactic anisotropy, we have to wait for a
definite answer for the first results of the Auger experiment.

{\em 4. LSP as UHE primary:\/}
An experimentally challenging but theoretically very clean signal both
for supersymmetry and for top-down models would be the detection of
the LSP as an UHE primary~\cite{Berezinsky:1997sb,Berezinsky:yk}.   
A decaying supermassive $X$ particle initiates a particle cascade
consisting mainly of gluons and light quarks but also of gluinos,
squarks and even only electroweakly interacting particles for
virtualities $Q^2\gg m_W^2, M_{\rm SUSY}^2$. When $Q^2$ reaches
$M_{\rm SUSY}^2$, the probability for further branching of
the supersymmetric particles goes to zero and their decays
produce eventually UHE LSPs. Signatures of UHE LPSs are 
a Glashow-like resonance at $10^9$~GeV $M_e/$TeV, where $M_e$ is the
selectron mass, and up-going showers for energies where the Earth is
opaque to neutrinos~\cite{Berezinsky:1997sb,BDHH}.

{\em 5. Clustering:\/}
Clustering of UHECR arrival directions could be explained in the
SHDM model by the clumpiness of the DM~\cite{Bl00a}. Although a
clumpy substructure of CDM is found both in analytical calculations and
numerical simulations, it is currently very uncertain how strong CDM is
clumped. Therefore clustering is difficult to use at present as an 
experimental constraint for SHDM. Any analysis should take into
account the spectral shape of the UHECR flux predicted in the SHDM model,
e.g., by using only events above $E>6$--$8\times 10^{19}$~eV.

The signatures of TD models are not so clear-cut, especially if TDs
contribute only a minor part to the UHECR flux.
The high photon/nucleon ratio at generation can be masked by
strong absorption of UHE photons, but is still higher than
expected from astrophysical sources, cf.~Fig~\ref{ABK}.a. 
All TD models predict 
large fluxes of UHE neutrinos. The GZK cutoff is less
pronounced for TDs than for astrophysical sources, because of the
flatter generation spectrum of the UHE particles. 
No clustering is expected in TD models because TDs emit UHE particles
in singular events.  
Finally, the detection of UHE LSPs is simpler in TD models than for
SHDM, because the predicted event numbers are higher for the same
UHECR flux.

\section{New primaries}
\label{newp}

Any new primary invented to explain the observed UHECR events needs a 
cross section with nucleons close to the ones typical for hadrons  and
a large energy transfer in each interaction in order to mimic the
observed properties of EAS. This requires a rather light particle with
strong or at least electromagnetic interactions, as one can see for a
composite particle $Y$ like a hadron or magnetic monopole from
kinematical considerations~\cite{Berezinsky:1997sb}: 
Such particles can have large total cross sections with nucleons even
when they are heavy, because they contain light constituents as 
e.g. gluons. However, the momentum fraction carried by the light
constituents goes to zero for $M_Y\to\infty$, and therefore also the
energy transferred in soft interactions. Moreover, the relative weight of
hard interactions in which a large longitudinal momentum transfer is
strongly suppressed by kinematics increases for large $M_Y$. As a
result, heavy hadrons or magnetic monopoles behave as deeply
penetrating particles although they have large total cross sections.
This behaviour can be seen nicely in Fig.~\ref{fit}.b where the
longitudinal shower profiles for EAS of energy $E_0=3\times 10^{20}$~eV
initiated by glueballinos---a bound-state of a gluon and gluino---with 
mass $M_{\tilde g}=2,5,10$ and 50~GeV
are compared to the EAS initiated by a proton. In this example, the
total cross section decreases only from 
$\sim 100$~mbarn ($M_{\tilde g}=2$~GeV) to
$\sim 90$~mbarn ($M_{\tilde g}=50$~GeV), while the energy fraction
transferred reduces from $y\sim 0.25$ to $0.02$,
cf. Ref.~\cite{Berezinsky:2001fy}.

On the other hand, the GZK cutoff for the new primary should be
shifted at least to $\gsim 10^{20}$~eV. This can be achieved by
requiring that the new primary is heavier than a nucleon.
Combining these two requirements, Ref.~\cite{Berezinsky:2001fy} found
that a viable new hadron should have a mass in
the range 2~GeV$\lsim m \lsim
5$~GeV. Reference~\cite{Kachelriess:2003yy} discussed the question if
such particles can be produced in astrophysical sources without
violating bounds like the EGRET limit. The authors concluded the
production of a new hadronic primary is only possible in collisions on
background photons and for masses smaller than $\lsim 3$~GeV.
Thus there is in principle a mass window around 2--3 GeV where a new
hadron could be a viable UHECR primary. But is there any candidate for
such a light hadron and a life-time above $\sim 1$~year, needed to
survive its journey? 

Until recently, the most discussed possibility of this kind was a
gluino as the LSP 
or next-to-LSP. However, measurements of electroweak observables
at LEPI were used in Ref.~\cite{Janot:2003cr} to constrain production
processes of new particles, and a light gluino with mass below
6.3~GeV was excluded at 95\% CL. The only remaining possibility in the
minimal supersymmetric SM for a strongly interacting LSP is a light
sbottom quark. It is likely that the lightest stable hadron
is charged in this case and thus it is implausible that it  remained
undetected in (accelerator) experiments, if it is stable.

A similar argumentation can be used against other, non-hadronic
primaries. 
Both the characteristics of EAS and the requirement of efficient
production in astrophysical beam-dumps require rather large couplings
of any primary to nucleons and photons. Together with the
bound on its lifetime, $\gsim 1$~year, this makes it rather
implausible that such a particle has not been detected yet.
Conceptionally different is the possibility that new particles are not
produced via a beam-dump, but are either accelerated directly in the
source or are produced during the propagation via mixing. A model with an
axion-like particle, i.e. a scalar which can mix with a photon in the
presence of 
external magnetic fields, was suggested in  Ref.~\cite{axion}. 
Axion-like particles can be also produced by photons emitted by 
astrophysical sources  via axion-photon oscillations~\cite{Csaki:2003ef}. 
The main problem of this type of model is the EGRET bound.
Bound-states of magnetic monopoles as primaries were proposed in
\cite{Huguet:1999bu}. From the discussion above, it is clear that
although their total cross section with nucleons can be large, the
energy transfer per interaction is very small. Thus they would behave as
deeply penetrating particles and cannot explain the observed EAS.

\section{Violation of Lorentz invariance}
Planck introduced already more than 100 years ago as fundamental length
scale $\ell_{\rm Pl}\equiv \sqrt{hG/c^3}\sim 10^{-33}$~cm. Today, it is
still an open question if $\ell_{\rm Pl}$ plays just the role of a
dimensionful coupling constant for gravity or if for smaller (wave-)
lengths  the properties of space-time are changed. If one considers,
e.g., the case that $l_{\rm Pl}$ sets a minimum wavelength in a 
frame-independent way, then it is clear that
special relativity has to be modified: Lorentz symmetry has to be either   
broken (a preferred inertial system exists) or ``deformed.'' In the latter
case, the usual Lorentz transformations are the limit $l_{\rm Pl}\to
0$ of more general transformations, similar as Galilei transformations
are obtained in the limit $c\to\infty$ from Lorentz transformations.
Other schemes in which modifications of Lorentz invariance are expected
in a purely four-dimensional frame-work are discrete (e.g. from
loop quantum gravity) or noncommutative space-times. 
Yet another possibility is that in topologically non-trivial space-times, 
as suggested by ``space-time foam'' \`a la Wheeler, chiral gauge theories
have a CPT anomaly which induces violation of Lorentz invariance~\cite{cpt}. 
Finally, Lorentz invariance could be violated only from our
(3+1)-dimensional point of view, while the underlying
higher-dimensional theory respects Lorentz symmetry.  In this scheme, the
slightly different localization of various SM particles on our
(3+1)-dimensional brane would induce modifications of Lorentz invariance.

Lorentz invariance violation can be implemented in an effective way by
allowing different maximal velocities for different particle
species~\cite{Coleman:1998ti}. 
The two most important consequences are changed dispersion
relations~\cite{Pavlopoulos:1967dm}, 
e.g. an energy dependent speed $v$ of (nearly) massless particles like
photons and neutrinos, and changed kinematical thresholds in
scattering and decay processes. For signals with a very
short duration and at cosmological distance like gamma-ray-bursts, the
energy dependence of $v$ could result in a detectable shift in the
arrival time of specific burst patterns at different
frequencies~\cite{grb}.

The change of kinematical thresholds in scattering processes could have a
dramatic impact on UHECRs if the threshold of the GZK reaction
$p+\gamma_{3K}\to N+\pi$ would be shifted to higher
energies~\cite{kc}. Apart from the extension of the UHECR spectrum beyond
$E_{\rm GZK}$,  the non-observation of GZK neutrinos would be 
characteristic for this solution to the UHECR puzzle. 
Moreover, Ref.~\cite{Dubovsky:2001hj} suggested as additional, but
model-dependent 
signature two sharp transitions in the composition of UHECRs:
Above a certain threshold energy $E_1$, neutrons become stable and
protons as primaries would be replaced by a neutron/proton mixture. 
Above a second threshold $E_2>E_1$, only neutrons
would be UHECR primaries.    
The reason for this mutation of the primary composition is the changed
dispersion relation of nucleons that above $E_1$  prohibits normal beta
decay and above $E_2>E_1$ allows the inverse beta decay $p\to
n+e^++\nu_e$. This change in the UHECR composition could be detected
via  a (non-) deflection of the neutron/proton primaries in the
galactic magnetic field, if the UHECRs correlate with astrophysical
sources.

\section{Conclusions}
Many explanations for the observation of UHECRs beyond the GZK cutoff
have been proposed during the last two decades that involve particle
physics beyond the standard model. The degree to which they solve
the difficulties of the conventional, ``bottom-up''  scenario is very
different: while the $Z$ burst model even aggravates the acceleration
problem, top-down models circumvent this issue by construction and,
in the particular case of SHDM, predict even no GZK cutoff at all.  
The latter can also be the case if Lorentz invariance is violated.

The combination of results from low-energy gamma-ray and UHE neutrino
experiments severely constrains TD and $Z$ burst models already now.
In the near future, the Pierre Auger Observatory will not only
answer the question  up to which energies the UHECR energy spectrum
extends, but also check conclusively the two key
signatures of SHDM, galactic anisotropy and photon dominance. 

Lorentz invariance violation should be considered seriously as an
explanation for the UHECR puzzle, if the spectrum extends well beyond
the GZK cutoff, there is not a considerable fraction of photon
primaries at the highest energies, and correlations with sources at
cosmological distance can be established.

\Acknowledgements{I am grateful to all my co-authors but in particular
to  Venya 
Berezinsky, Sergey Ostapchenko, and Dima Semikoz for fruitful
collaborations and many discussions. I would like to thank
Andreas Ringwald, G\"unter Sigl and Peter Tinyakov for helpful comments. 
This work was supported by the Deutsche Forschungsgemeinschaft
within the Emmy-Noether program.}

%
\end{document}